# Can experiments determine the stacking fault energies of metastable alloys?

Xun Sun[a,b], Song Lu[a,*], Ruiwen Xie[a], Xianghai An[c], Wei Li[a], Tianlong Zhang[b], Chuanxin Liang[b], Xiangdong Ding[b], Yunzhi Wang[d], Hualei Zhang[b,**] and Levente Vitos[a,e,f]

[a]Applied Materials Physics, Department of Materials Science and Engineering, Royal Institute of Technology, Stockholm SE-100 44, Sweden

[b]Frontier Institute of Science and Technology, State Key Laboratory for Mechanical Behavior of Materials, Xi'an Jiaotong University, Xi'an, 710049, China

[c]School of Aerospace, Mechanical & Mechatronic Engineering, The University of Sydney, Sydney, NSW 2006, Australia

[d]Department of Materials Science and Engineering, The Ohio State University, 2041 College Road, Columbus, OH 43210, USA

[e]Division of Materials Theory, Department of Physics and Materials Science, Uppsala University, P.O. Box 516, SE-75120 Uppsala, Sweden

[f]Research Institute for Solid State Physics and Optics, Wigner Research Center for Physics, Budapest H-1525, P.O. Box 49, Hungary

Corresponding authors: songlu@kth.se (S.L.) and hualei@xjtu.edu.cn (H.Z.)

## Abstract:

Stacking fault energy (SFE) plays an important role in deformation mechanisms and mechanical properties of face-centered cubic (fcc) metals and alloys. In metastable fcc alloys, the SFEs determined from density functional theory (DFT) calculations and experimental methods often have opposite signs. Here, we show that the negative SFE by DFT reflects the thermodynamic instability of the fcc phase relative to the hexagonal close-packed one; while the experimentally determined SFEs are restricted to be positive by the models behind the indirect measurements. We argue that the common models underlying the experimental measurements of SFE fail in metastable alloys. In various concentrated solid solutions, we demonstrate that the SFEs obtained by DFT calculations correlate well with the primary deformation mechanisms observed experimentally, showing a better resolution than the experimentally measured SFEs. Furthermore, we believe that the negative SFE is important for understanding the abnormal behaviors of partial dislocations in metastable alloys under deformation. The present work advances the fundamental understanding of SFE and its relation to plastic deformations, and sheds light on future alloy design by physical metallurgy.

## 1. Introduction

Intrinsic stacking fault (ISF) in face-centered-cubic (fcc) materials is a type of two-dimensional defect which is usually created by the glide of a Shockley partial dislocation during deformation. Immediately, it is related to the nucleation of deformation twins and hexagonal close-packed (hcp) martensite which are generated by collective motions of partials. ISF is also the characteristic structure between the trailing and leading partials in a dissociated $a/2<110>$ dislocation. From these aspects, the excess formation energy of ISF, i.e., the stacking fault energy (SFE), serves as a significant intrinsic material parameter affecting the dissociation of $a/2<110>$ dislocations, twinning as well as martensitic transformation (MT). For example, in pure fcc metals, deformation twinning (DT) is often observed in Ag with a low SFE (measured value 16±2 mJ/m$^2$ [1]) but rarely in Al with a very high SFE (measured values ~150±40 mJ/m$^2$ [2]). In austenitic Fe-Cr-Ni and Fe-Mn steels [3-9], an empirical relation between the prevalent deformation mode and the size of the measured SFE is established, showing that in addition to dislocation slips, low-SFE alloys (<~20 mJ/m$^2$) prefer deformation-induced martensitic transformations (DIMTs) from fcc ($\gamma$) to hcp ($\varepsilon$) or to body-centered-tetragonal (bct, $\alpha'$) phases, while medium-SFEs (~20-40 mJ/m$^2$) prefer DT [3, 6, 8, 10, 11]. In materials with high SFEs, usually only dislocation slips can be observed under normal loading conditions [11]. The occurrence of DT or MT during plastic deformation provides important internal boundaries that interact with dislocations in distinct ways from grain boundaries [12], which enables the so-called dynamical Hall-Petch effect, maintaining the high work-hardening rate and thereby delaying the onset of plastic instability and necking [13, 14]. The resulting twinning-induced plasticity (TWIP) and transformation-induced plasticity (TRIP) are two important mechanisms responsible for the excellent plastic properties of high-Mn austenitic steels [3, 4, 6-8] as well as some multicomponent solid solutions (also called medium- or high-entropy alloys, MEA/HEA) such as CrCoNi and CrMnFeCoNi [15-20]. Therefore, in the development of high-strength alloys, significant effort has been put to study factors that may affect the SFE, such as temperature [9, 21-24], composition [10, 21, 25] and short-range order [26-31]; to understand the microscopic mechanisms underlying the role of SFE in DT, MT and dislocation planar slip [4, 7, 10, 32-35]; as well as to quantitatively predict the critical twinning stress based on the SFE [3, 18, 34, 36-41] or the generalized SFE (γ-surface) [42-44]. All these activities depend on the accurate determination of the SFE.

Nowadays, both experimental [45-47] and computational methods [42, 48-52] are commonly applied to evaluate the SFEs. Particularly, ab initio methods based on density functional theory (DFT) calculations have been widely adopted to determine the SFE at 0K, as well as at elevated temperatures via considering contributions from thermal lattice expansion and electronic/phononic/magnetic excitations [23, 53-56]. For pure fcc metals, a satisfying agreement between experimental and ab initio results has been reached for both the SFE values and the variation trends with respect to temperature [48, 53, 54]. However, the situation becomes more complicated for some alloys, especially the concentrated ones. For example, in Cu-Al alloys ab initio calculations showed that the SFE decreases approximately linearly with increasing Al concentration and turns to negative at ~10 at.% Al [48, 57]; whereas the experimentally determined values first decrease and then become constant at ~5 mJ/m$^2$ for Al concentration higher than 10 at.% [22]. Similar observation can be made for Ni-Co alloys with respect to Co concentration [21, 22, 58-60]. It seems that the temperature effect cannot systematically improve the agreement between the theoretical calculations and experimental measurements of SFEs in pure metals (/dilute alloys) and concentrated alloys. Recently, in the development of HEAs, the experimental and theoretical SFEs often have opposite signs (see Table 1), which cannot be ascribed solely to the inappropriate treatment of the temperature effect in ab initio calculations [13, 15, 17, 26, 61-68]. Further efforts to resolve the problem include the consideration of short-range order (SRO) or local chemical variations. For instance, atomistic studies demonstrated that the SFE depends on the local chemical composition as well as chemical and magnetic SROs [26-31]. Ding et al. [26] reported that the calculated mean SFE of CrCoNi MEA increases markedly from -42.9 mJ/m$^2$ at the fully random state (estimated through averaging the SFEs of 108 configurations with a broad distribution from -140 to 65 mJ/m$^2$) to ~30 mJ/m$^2$ with increasing the degree of SRO, in comparison to the experimental values, 22±4 [61] and 18±4 mJ/m$^2$ [17]. Thereby, the authors ascribed the discrepancy in the SFEs in these multicomponent alloys to the SROs. Li et al. [28] further argued that the wide variety of local chemical ordering produces a wide range of generalized stacking fault energies, therefore, increasing the ruggedness of the energy landscape for dislocation activities and influencing the selection of dislocation pathways in slip, faulting and twinning. Being aware of the difficulties of experimental methods in characterizing SROs, some recent experiments have found that in the homogenized CrCoNi specimens, SRO (or local chemical variation) is however very weak, if not absent [69]; only by aging the homogenized CrCoNi MEA at 1000 °C for 120 h followed by furnace cooling, Zhang et al. [62] reported observation of SRO domains of ~1 nm and that the measured SFE based on partial separation distance increases from 8.18±1.43 for the water-quenched sample to 23.33±4.31 mJ/m$^2$ for the annealed one. Therefore, the discrepancy in the theoretical and experimental SFEs is not solved in CrCoNi MEA, neither in other HEAs. Additionally, one may notice the different experimental SFE values for the homogenized CrCoNi MEA in different Refs.[17, 61, 62], despite that they are all based on transmission electron microscopy (TEM) measurements of partial separation widths, which was also ascribed to the inherent local chemical variations in these alloys [29]. Smith et al. [29] found large variations in dislocation dissociation distances through analyzing the measured results for 30 different $a$/2<110>{111} 60° mixed dislocations in the CrMnFeCoNi HEA. The average separation distance was 4.82 nm (~5.5 nm [13]) but with a large range of variation (±3.4 nm), an order of magnitude higher than the normal variation (±0.45 nm) in pure fcc metals [29]. No evidence of alloying element segregation or ordering was found in the alloy [70], but the large variations in dislocation separations were still ascribed to the local chemical inhomogeneity. It is however not clear how the local composition inhomogeneity or SROs, if exist, in the length scales of ~1 nm or less [26, 62, 69] affect the behaviors of dislocations with significantly longer lengths and larger separation widths in the processes of faulting or twinning.

Ab initio calculations for the thermodynamic parameters like the SFE as a function of composition in alloys are efficient in screening for proper candidates before conducting experiments. In order to facilitate future design of alloys with vast composition space, it is important to better understand the ab initio and experimental methodologies for the SFE determination, especially their limitations. Here, we aim at answering the following questions: can experimental methods determine the SFEs of metastable alloys and what is the relationship between the measured and the ab initio SFEs in these systems? In the present work, we focus on the metastable alloys defined by a higher Gibbs free energy of the fcc phase than that of the hcp one ($\Delta G^{\gamma \to \varepsilon} < 0$), which is most relevant case for the discussions of the deformation mechanisms in fcc solid solutions. Although metastability may also refer to the thermodynamic states of other phases such as decomposed phases and precipitates, their impact is not discussed here.

The rest of the paper is organized as following. In Section 2, we give a brief introduction of the adopted theoretical approaches for the SFE calculations, more details of the methodologies can be found in our previous publications [56, 71]. Results and discussion are given in Section 3. We start with analyzing the limitations of current experimental approaches for measuring the SFE in subsection 3.1. The correlations between the calculated/experimental SFEs and phase stability are presented in subsection 3.2. In subsection 3.3 we analyze the correlation between the theoretical SFEs and the observed deformation mechanisms. Further implications of the present findings are discussed in subsection 3.4. Conclusions are drawn in Section 4.

## 2. Methodology

All DFT calculations were performed using the exact muffin-tin orbitals (EMTO) method [72] in combination with the single-site coherent potential approximation (CPA) [73]. Both the self-consistent calculations and total energies were calculated within the Perdew-Burke-Ernzerhof (PBE) exchange-correlation functional [74]. The scalar-relativistic approximation and soft-core scheme were adopted to solve the one-electron Kohn-Sham equations. The Green's function was calculated self-consistently for 16 complex energy points. We employed an $l$-cutoff of 8 in the one-center expansion of the full-charge density. The paramagnetic (PM) state was described by the disordered local magnetic moment (DLM) model [75, 76]. The 9×18×2 k-point mesh was used for all supercell calculations after careful test. A careful assessment of the present method for the determination of the SFE in fcc metals and alloys can be found in Ref. [48]. The SFEs were calculated using supercells formed by 12 close-packed {111} layers. The intrinsic stacking fault (ABC|BCABC) is formed when a layer of atoms is removed from a perfect fcc sequence (ABCABCABC) [47]. The SFE was computed as $SFE = (F^{\text{fault}} - F^0)/A$, where $F^{\text{fault}}$ and $F^0$ are the free energies with and without a stacking fault, respectively, $A$ is the stacking fault area. In the previous works [42, 48, 56], this supercell method was successfully used to calculate the SFEs of pure metals and solid solutions. The free energies were approximated as $F=E-TS_{\text{mag}}$, where $T$ is the temperature, $E$ is the internal energy calculated at 0K. The magnetic entropy $S_{mag} = k_B \sum_{i=1}^{n} c_i \ln(1 + \mu_i)$ was calculated within the mean-field approximation, where $k_B$ is the Boltzmann constant, $n$ is the total number of elements, $c_i$ and $\mu_i$ are the concentration and local magnetic moment for the $i$th alloying element, respectively. To partly include the room temperature (RT) effect on the SFE, we used the experimental lattice parameters measured [61, 64, 77, 78] at RT or obtained by a regression formula [77]. Electronic and explicitly anharmonic phonon contributions were estimated to be small at RT and they were neglected [79], which however does not affect the discussion in the present work. For more details regarding the treatment of the temperature effect on the SFE calculation, readers are referred to Refs. [53, 54, 80, 81].

## 3. Results and discussion

### 3.1 Limitations of experimental methods for SFE determination

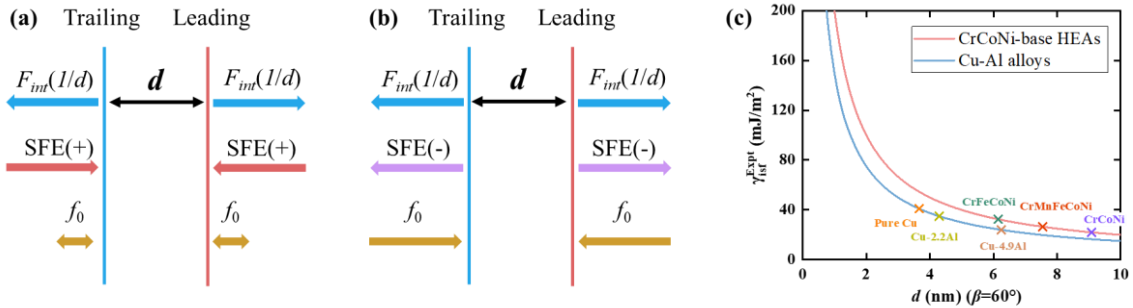

**Figure 1.** Schematic of the force balance for partial dislocations in materials with positive SFE (a) and with negative SFE (b) at static conditions. $F_{int}$ stands for the elastic repulsive force (per unit length) between two partials and is inversely proportional to the partial separation with $d$. Although the lattice friction force $f_0$ is present in both cases, its role becomes critical in negative-SFE materials. In the case of positive SFE, at static conditions, the direction of the passive $f_0$ depends on the difference between SFE and $F_{int}$ [82]. In the case of negative SFE in (b), $f_0$ must be opposite to the direction of $F_{int}$ and the SFE, resisting dislocation dissociation. (c) The inverse proportion relationship between $\gamma_{isf}^{Expt.}$ and *d*. Examples are given for Cu-Al alloys [1, 83] and CrCoNi-based HEAs [15, 17, 61] with different shear moduli. The marked partial separation widths were measured for dislocations with $\beta = 60°$.

SFE is usually considered as an experimentally accessible parameter by means of TEM, X-ray or neutron diffractions. In the following, we discuss the limitations of the experimental methods for determining the SFE, especially in metastable alloys. Methods involving TEM measure the distance (*d*) between the two *a*/6<112> partial dislocations in a dissociated *a*/2<110> dislocation [45] or the radii describing the size of the dislocation node [46], and then one connects these parameters to the SFE. In order to minimize the interaction between dislocations, the partial separation width *d* is usually measured on an isolated straight dislocation line at its equilibrium state, then the SFE is calculated according to

$$\gamma_{isf}^{Expt.} = f(b_p, \mu, \nu, \beta)/d \quad \text{with} \quad f(b_p, \mu, \nu, \beta) = \frac{\mu b_p^2}{8\pi} \frac{2-\nu}{1-\nu}\left(1 - \frac{2\nu \cos 2\beta}{2-\nu}\right), \quad (1)$$

where $\mu$ is the shear modulus on the {111} close-packed planes, $b_p$ is the Burgers vector of partial dislocation, $\nu$ is the Poisson ratio, $\beta$ is the angle between Burgers vector *b* and dislocation line direction. *d* may also be called the stacking fault width (SFW). It is important to notice that the above equation is established based on the presumed balance between two forces acting on a partial dislocation: namely the repulsive force ($F_{int}$) (per unit length of dislocation line) due to the elastic interaction between the two partials (leading and trailing) and the attractive force due to the excess energy cost to create a stacking fault (SF), i.e., the SFE (Figure 1(a)). The right hand term in Eq. (1) corresponds to the repulsive force calculated according to the classic theory of dislocation [47]. In other words, $\gamma_{isf}^{Expt.}$ is indirectly obtained from the elastic repulsive force through the measured $d$. Several critical points can then be drawn from Eq. (1).

(1) For dislocations of the same character ($\beta$), the $F_{int}$ and the so-calculated $\gamma_{isf}^{Expt.}$ are inversely proportional to $d$, which are always positive and approach to zero as $d$ increases, see Figure 1(c).

(2) In the case of small partial separation (corresponding to large SFE), a small scatter in $d$ measurement can cause a large deviation in $\gamma_{isf}^{Expt.}$, thereby, the TEM method is usually thought to be proper for measuring small SFEs to avoid the large uncertainty [1, 17, 61, 64].

(3) In the case of large separation, the $\gamma_{isf}^{Expt.}$ quickly decreases toward a small positive value (e.g., < 20 mJ/m$^2$) which barely changes with increasing $d$ within the experimental error bar; in other words, $\gamma_{isf}^{Expt.}$ gradually loses the resolution in identifying the dislocation separation. This observation is important for understanding the temperature effect on the experimental SFE in various alloys. For example, for alloys with positive temperature dependence of the SFE, the measured SFE appears almost constant at low temperatures [84-87].

(4) For fcc metals and alloys, $\nu$ is usually around 0.3 [13, 17]. The proportional

factor ($f(b_p, \mu, \nu, \beta)$) mainly depends on the corresponding shear moduli of different materials. For the same partial separation width, the $\gamma_{isf}^{Expt.}$ can differ significantly in alloys with different shear moduli, which explains why the critical $\gamma_{isf}^{Expt.}$s separating different deformation mechanisms differ significantly in various alloys [11]. In other words, the same $\gamma_{isf}^{Expt.}$ value in different alloy systems may correspond to different partial separation widths and indicates different states of thermodynamic stability of the fcc phase.

Most critically, the true SFE has always to be positive to potentially allow Eq. (1) to hold, neglecting any other factors (e.g., the lattice friction force [34, 82]) that can participate in establishing the equilibrium stacking fault ribbon. On the other hand, the sign of the SFE can to some extend indicate the thermodynamic stability of the fcc phase as discussed in the following. First, by definition, the SFE is the excess formation energy of the intrinsic stacking fault relative to the energy state of the fcc structure. From the structural point of view, the stacking fault can be seen as a two-layer of hcp embryo embedded in the fcc matrix, thereby the SFE can be expressed as [5]

$$\text{SFE} = \frac{2\,(\Delta G^{\gamma \to \varepsilon} + E^{str.})}{A} + 2\sigma^{\varepsilon/\gamma}, \tag{2}$$

where $\Delta G^{\gamma \to \varepsilon}$ is the Gibbs energy difference (per atom) between the hcp and fcc phases, $E^{str.}$ is the strain energy contribution, $A$ is the stacking fault area and $\sigma^{\varepsilon/\gamma}$ is the coherent interfacial energy between fcc and the two-layer hcp embryo [5]. $E^{str.}$ is a small positive term giving a contribution of ~1-4 mJ/m$^2$ [5, 88]. The interfacial energy for the coherent fcc{111}/hcp(0001) interface depends on the thickness of the hcp phase;

and therefore $\sigma^{\varepsilon/\gamma}$ for the two-layer hcp embryo usually differs from the value obtained between bulk fcc and bulk hcp [89]. Previous ab initio studies showed that $\sigma^{\varepsilon/\gamma}$ is in the range of ±9 mJ/m² [89]. When the Gibbs energy difference is small, $\sigma^{\varepsilon/\gamma}$ and $E^{str.}$ together determine the sign of the SFE. However, when the hcp structure is much more stable than the fcc one (typically, $\Delta G^{\gamma \to \varepsilon} \lesssim -0.2$ mRy/atom), the SFE of the metastable fcc phase as computed from Eq. (2) becomes negative. Obviously in this case one cannot use Eq. (1) to determine the SFE because both the SFE and the elastic repulsive force point to the same direction and the force balance condition between these two terms breaks down (Figure 1(b)). We argue that in the absence of external shear stress, it is the lattice friction force on the partial dislocations which can ensure the force balance condition and thus lead to finite partial separations in metastable systems [34, 82]. Therefore, for metastable fcc alloys, Eq. (1) breaks down and cannot be adopted for deriving the experimental SFE. We note that in the positive-SFE cases like pure fcc metals, the lattice friction force/stress should also exist, but it does not play a dramatic role as in the case of negative SFE, which will be further discussed in the following sections.

Similar arguments apply for the SFE measurement by using the dislocation node method [46, 90]. Furthermore, measuring the SFE by using X-ray or neutron diffraction suffers from similar restrictions since the support for the measurement is also based on the above-discussed force balance assumption to derive the SFE [91]. There, the partial separation width $d$ is replaced by the stacking fault probability through $\alpha = \frac{\rho d a_0}{\sqrt{3}}$, where $a_0$ is the lattice constant, and $\rho = K_{111}\langle\epsilon_{50}^2\rangle_{111}/b^2$, $K_{111}$ is a constant and $\langle\epsilon_{50}^2\rangle_{111}$ is the rms microstrain in the [111] direction averaged over a distance of 50 Å. $\alpha$ is then measured from the shift in the position of the diffraction lines [92]. Consequently, $\gamma_{isf}^{Expt.}$ obtained by those experimental methods will all depart from the thermodynamic stability of the fcc phase (expressed via $2\Delta G^{\gamma\to\varepsilon}/A$) in metastable systems.

## 3.2 Correlations between ab initio SFE, $\Delta G^{\gamma\to\varepsilon}$, and experimental SFE

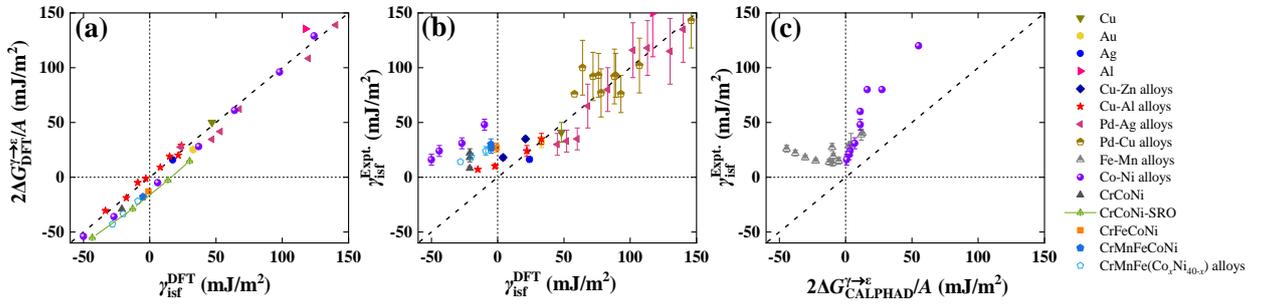

**Figure 2.** (a) The $\gamma_{isf}^{DFT}$ versus $2\Delta G^{\gamma\to\varepsilon}/A$ relationship from DFT calculations. (b) The $\gamma_{isf}^{DFT}$ versus $\gamma_{isf}^{Expt.}$ correlation. (c) The $\gamma_{isf}^{Expt.}$ versus $2\Delta G^{\gamma\to\varepsilon}/A$ relationship, where $\Delta G^{\gamma\to\varepsilon}$ data is from CALPHAD calculations. The experimental data is taken from Refs. [1, 2, 13, 17, 21, 22, 45, 61, 62, 64, 83, 88, 93-95] and the DFT results are from present work and from Refs. [26, 48, 89, 96]. CALPHAD results for Fe-Mn alloys are from Ref. [88], and those for Co-Ni alloys are calculated using TTNI8 database [97].

In Figure 2(a), we compare $\gamma_{isf}^{DFT}$ and $2\Delta G^{\gamma\to\varepsilon}/A$, both calculated by DFT method for various fcc metals and alloys. The two quantities correlate nicely with each other, as described in Eq. (2), even when SROs are present (points connected by lines) [26]. The observation is consistent with the previous studies [26, 48, 96, 98]. The strain energy contribution is not included in the calculations, therefore, the deviation between the two quantities gives the interfacial energy term ($2\sigma^{\varepsilon/\gamma}$). $\sigma^{\varepsilon/\gamma}$ is estimated in the range of -9~8 mJ/m$^2$ for the studied metals and alloys shown in Figure 2(a), which is in agreement with the previous values calculated by DFT methods [89].

In Figure 2(b), we compare the SFEs obtained by DFT ($\gamma_{isf}^{DFT}$) and by experiments ($\gamma_{isf}^{Expt.}$) for various thermodynamically stable and unstable fcc metals and alloys. One can observe that for fcc elemental metals (Ag, Cu, Au and Al, etc.) and the stable fcc alloys (e.g., Pd-Ag and Pd-Cu alloys) with $\Delta G^{\gamma \rightarrow \varepsilon} > 0$ (Figure 2(a)), DFT and experiments lead to consistent results, implying that the force balance condition assumed in Eq. (1) is indeed satisfied (Figure 1(a)), in other words, the lattice friction forces are negligible in these cases. We may estimate the magnitude of the lattice friction force from the Peierls stress ($\tau_P$) in pure fcc metals. Experimentally, the upper limit of Peierls stress is usually estimated from the critical resolved stress for slip extrapolated to 0 K, which is less than ~10 MPa for pure fcc metals ($\tau_P/G$~10$^{-4}$-10$^{-5}$, $G$ is the shear modulus, see Refs. [99, 100] and Refs. therein). Then the lattice friction force can be calculated as $\tau_P \cdot b \approx 3$ mJ/m$^2$ (Peach-Kohler force [47]), which falls in the range of the discrepancy between $\gamma_{isf}^{DFT}$ and $\gamma_{isf}^{Expt.}$ for pure fcc metals.

But for those metastable fcc alloys ($\Delta G^{\gamma \rightarrow \varepsilon} < 0$) including binary Co-Ni alloys, Cu-Al alloys with high Al concentrations, ternary Cr-Co-Ni MEAs and quinary Cr-Mn-Fe-Co-Ni HEAs, the theoretical and experimental values deviate notably from each other, which is primarily ascribed to the reasons discussed in Section 3.1. The uncertainty in DFT calculations are discussed in the following, but they cannot be responsible for the opposite signs of $\gamma_{\text{isf}}^{\text{DFT}}$ and $\gamma_{\text{isf}}^{\text{Expt.}}$ in these metastable alloys, especially when $\Delta G^{\gamma \to \varepsilon}$ is very negative. Similar conclusion can be reached when comparing the experimental SFE and the Gibbs energy difference $2\Delta G^{\gamma \to \varepsilon}/A$ obtained by CALPHAD calculations (Figure 2(c)).

Since pairs of partial dislocations are still observed in these metastable alloys, one expects the presence of large friction forces preventing the separation of full dislocations into very wide SFs or the spontaneous formation of hcp martensite. For example, in the case of CrCoNi MEA, the Peierls stress was estimated to be ~160 MPa, approximately one order of magnitude higher than that of Ni (~15 MPa) [101]. Yoshida et al. [102] measured the friction stress term ($\sigma_0$) in the Hall-Petch relation, $\sigma_{YS} = \sigma_0 + k\lambda^{1/2}$, where $\sigma_{YS}$ is the yield strength, $k$ is the Hall-Petch slope and $\lambda$ is the mean grain size. They showed that the friction stress is significantly larger for CrCrNi (218 MPa) and CrMnFeCoNi (125 MPa) than those for pure Al (4 MPa) and Ni (14 MPa). The large friction stresses in these alloys are related to the significantly roughened Peierls potential surface due to e.g., local lattice distortions and chemical variations [28, 102, 103].

In Table 1, we summarize the calculated SFE values by various DFT methods for the CrCoNi-based MEAs and HEAs as well as the $Fe_{40}Mn_{40}Co_{10}Cr_{10}$ TWIP HEA [26, 66-68], whose signs are all in stark contrast with the experimental ones [13, 15, 17, 61-65]. In particular, for the CrCoNi MEA which has been extensively studied by various experimental and theoretical methods due to its excellent mechanical performance, all DFT calculations yield negative $\gamma_{\text{isf}}^{\text{DFT}}$ and $\Delta G^{\gamma \to \varepsilon}$ for the random solid solution state at both room and cryogenic temperatures. $\gamma_{\text{isf}}^{\text{DFT}}$ and $\Delta G^{\gamma \to \varepsilon}$ are in nice correlation with each other, even when short-range orders are considered [26], which follows the consideration from the structural point of view. Despite the fact that variations in chemical or magnetic SROs may significantly alter the calculated SFE [26, 28, 30], the positive $\gamma_{\text{isf}}^{\text{Expt.}}$ comes inherently from the experimental methodologies and does not necessarily relate to the state of SROs in the materials [104]. Furthermore, recent experiments showed that the element distribution in the homogenized CrCoNi MEA is in fact highly random and homogeneous [69]. The metastability of the random CrCoNi MEA, as well CrMnFeCoNi HEA, is evidenced by the high pressure experiments showing that the fcc→hcp phase transformation occurs during pressing at ambient temperature [105-107] and by the coexistence of the hcp and fcc phases in the epitaxial film [108]. As discussed above, it is likely the large lattice friction stress [101, 102] that prevents the spontaneous nucleation of hcp martensite in these alloys at room temperature, despite that the hcp structure is thermodynamically more stable than the fcc one.

**Table 1.** Comparison between experimental ($\gamma_{\text{isf}}^{\text{Expt.}}$, mJ/m$^2$) and DFT ($\gamma_{\text{isf}}^{\text{DFT}}$, mJ/m$^2$) SFEs for MEAs and HEAs at room temperature. The $\gamma_{\text{isf}}^{\text{Expt.}}$s calculated from TEM measurements of dislocation line or node and from X-ray or neutron diffraction measurements of stacking fault probability are indicated by SFW, Node, XRD and ND, respectively. DFT results are obtained for the random solid solutions which are modeled by the coherent potential approximation (CPA) method or the special quasirandom structures (SQS). The corresponding temperature and magnetic state (NM: nonmagnetic, PM: paramagnetic, FM: ferromagnetic) at which the calculations are performed are also indicated.

| Alloys (at. %) | $\gamma_{isf}^{Expt.}$ | Experiment | $\gamma_{isf}^{DFT}$ | Calculation |
|---|---|---|---|---|
| CrCoNi | 18±4 [17] | SFW | -21 | CPA, 300K, PM |
|  | 22±4 [61] | SFW | -43 [26] | SQS, 0K, NM |
|  | 8.18±1.43 [62] | SFW | -24 [66] | SQS, 0K, NM |
|  |  |  | -62 [67] | SQS, 0K, NM |
|  |  |  | -38 [68] | SQS, 0K, FM |
| CrFeCoNi | 27±4 [17] | SFW | -1 | CPA, 300K, PM |
|  | ~20-25 [63] | XRD | -23 [67] | SQS, 0K, FM |
|  | 32.5 [15] | ND |  |  |
| CrMnFeCoNi | 30±5 [13] | SFW | -5 | CPA, 300K, PM |
|  | 26.5±4.5 [17] | SFW | -54 [67] | SQS, 0K, FM |
|  | ~20-25 [63] | XRD | -31 [68] | SQS, 0K, FM |
| $Cr_{20}Mn_{20}Fe_{20}Co_{23}Ni_{17}$ | 24±4 [64] | SFW | -9 | CPA, 300K, PM |
| $Cr_{20}Mn_{20}Fe_{20}Co_{27}Ni_{13}$ | 19±3 [64] | SFW | -20 | CPA, 300K, PM |
| $Cr_{20}Mn_{20}Fe_{20}Co_{30}Ni_{10}$ | ~14 [64] | SFW | -28 | CPA, 300K, PM |
| $Fe_{40}Mn_{40}Co_{10}Cr_{10}$ | 13±4 [65] | Node | -3 | CPA, 300K, PM |

## 3.3 Correlation between ab initio SFE and deformation mechanism

In literature, the occurrence of DT or γ→ε DIMT is usually rationalized with the sizes of the experimental SFEs. Being aware of the limitations of the experimental methods in determining the SFE, instead, here we may establish the relationship between the ab initio SFE and the deformation mode. In Figure 3, we demonstrate the correlation between $\gamma_{isf}^{DFT}$ and the observed prevalent deformation mechanisms [61, 64, 109-111] in a group of MEAs and HEAs. We emphasize here that despite of the significant complexity of the compositions of these multicomponent alloys, the calculated $\gamma_{isf}^{DFT}$ acts as a reliable indicator for the activation of the primary deformation mechanisms. Compared to the experimental SFEs (usually available for TWIP, but not for TRIP alloys due to the fact that a large amount of wide SFs exist already at very small strains, which prevents meaningful measurements of partial separation width), the ab initio SFEs have a better resolution in indicating the twinnability or the tendency to DIMT. For all the studied alloys, the $\gamma_{isf}^{DFT}$s are negative at room temperature, suggesting the metastable nature of these alloys. There seems a critical value or transition region of $\gamma_{isf}^{DFT}$ at around -20

mJ/m$^2$ corresponding approximately to the case of CrCoNi MEA, that separates the observed primary deformation mechanisms and the operation of the TWIP and TRIP effects. We should mention here that although DT is the dominated mechanism in the CrCoNi MEA, small amounts of hcp martensite has indeed been observed during deformation [112-114]. Similarly for the $Cr_{20}Mn_{20}Fe_{20}Co_{30}Ni_{10}$ HEA in the TRIP region, whose $\gamma_{isf}^{DFT}$ is -28 mJ/m$^2$, locating close to the transition zone, it was observed that SFs and DT occur at small strains and hcp MT becomes dominant at large strains [64]. Thus, these MEAs and HEAs with $\gamma_{isf}^{DFT}$ in the transition zone are likely to show joint TWIP+TRIP effects, together with simultaneously enhanced synergy of strengthen and ductility [17, 61, 64, 109]. Accordingly, for the $Cr_{25}Fe_{25}Co_{35}Ni_{15}$ HEA, whose $\gamma_{isf}^{DFT}$ is calculated to be -26 mJ/m$^2$, both DT and MT are expected from the nice correlation in Figure 3, although only the MT was reported from the electron backscatter diffraction (EBSD) results [111], which is likely due to the limited resolution of the EBSD images.

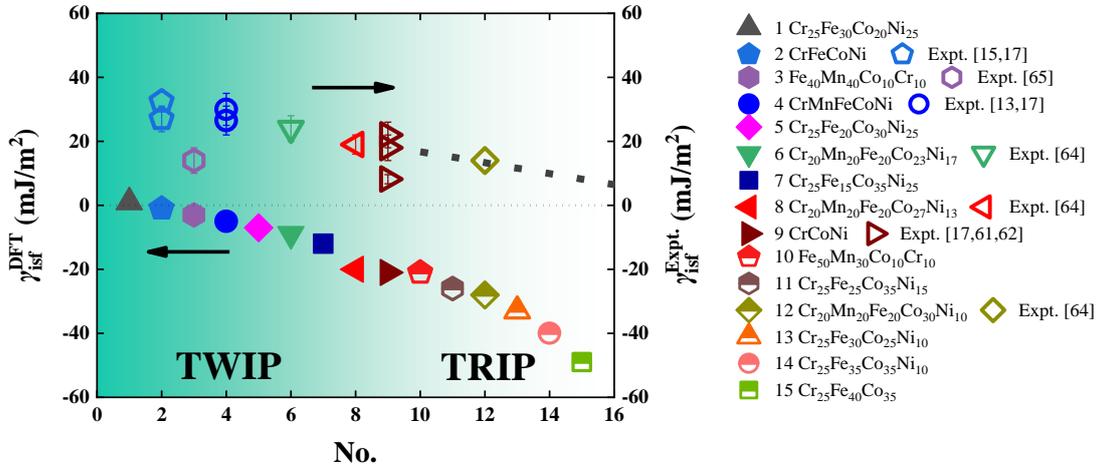

**Figure 3.** Correlation between the calculated SFEs ($\gamma_{isf}^{DFT}$, filled and half-filled symbols) and the observed strengthening mechanisms (TWIP/TRIP) [61, 64, 109-111] in a group of MEAs and HEAs. Available experimental SFEs ($\gamma_{isf}^{Expt.}$, empty symbols) are also plotted for comparison. Experimental SFEs for TRIP alloys are usually not available but should approach zero as indicated by the dashed line according to the experimental

formulas for SFE calculations. Consequently, even without considering the physical correctness of these $\gamma_{\text{isf}}^{\text{Expt.}}$ values, the ab initio SFEs have a better resolution than the experimental ones in indicating the twinnability or the tendency to DIMT.

### 3.4. Implications of negative SFE

In the following, we discuss some further implications of the negative SFE.

#### 3.4.1 Interfacial energy $\sigma^{\varepsilon/\gamma}$

One should be extremely cautious to determine the interfacial energy by subtracting the $\Delta G^{\gamma \to \varepsilon}$ contribution from the measured SFE [5, 88]. Since experimentally there is no direct method for measuring the interfacial energy, Olson and Cohen [5] originally proposed that the interfacial energy $\sigma^{\varepsilon/\gamma}$ can be determined via the measured SFE subtracted by the contribution from Gibbs free energy difference $\Delta G^{\gamma \to \varepsilon}$, i.e., $\sigma^{\varepsilon/\gamma} = \frac{\gamma_{\text{isf}}^{\text{Expt.}}}{2} - \frac{\Delta G^{\gamma \to \varepsilon}}{A}$ (from Eq. (2)). The $\Delta G^{\gamma \to \varepsilon}$s of Fe-Cr-Ni alloys were calculated as a function of temperature by using of the regular solution thermodynamics [115]. The $\gamma_{\text{isf}}^{\text{Expt.}}$s of Fe-Cr-Ni alloys were measured by the extended node method [5, 86, 116] at different temperatures. The interfacial energies $\sigma^{\varepsilon/\gamma}$ s of Fe-Cr-Ni alloys were accordingly determined to be ~10-15 mJ/m$^2$, which were later on widely adopted in the SFE calculations using the thermodynamic approaches [5]. In light of our above discussions, since the measured SFE decreases towards a small positive value with decreasing $\Delta G^{\gamma \to \varepsilon}$, the obtained interfacial energy in this way will increase by the similar magnitude as the decrease in the $\Delta G^{\gamma \to \varepsilon}$ (Figure 2(c)), as obtained in Ref.[88], which is an artificial effect. The obtained large interfacial energy does not describe the real energetics of the hcp/fcc interface. There are efforts to calculate the composition dependent SFE based on thermodynamic descriptions of $\Delta G^{\gamma \to \varepsilon}$ for the purpose to design TRIP/TWIP alloys, while the interfacial energy term is commonly treated as a fitting parameter to realize the agreement between the calculated SFE and the $\gamma_{\text{isf}}^{\text{Expt.}}$ [50, 51, 88, 117]. In these cases, the $\gamma_{\text{isf}}^{\text{Expt.}}$ values adopted to derive the interfacial energy become critical. A large interfacial energy can effectively wash out the important information carried by $\Delta G^{\gamma \to \varepsilon}$.

**3.4.2. The effect of negative SFE on the mobility of partial dislocations with respect to altering temperature or stress**

Negative SFE is important for understanding the temperature effect on the partial separation width, for the purpose to determine the temperature effect on the SFE [84, 94, 118]. Usually in pure fcc metals and dilute alloys, the temperature induced variation of the SFW is fully reversible. But in concentrated alloys such as in Ag-Al [118], Cu-Al [94], Co-Ni [84] and Co-Cr-Ni [84] alloys, below a certain temperature $T_0$ the SFW does not change with temperature. Above $T_0$, partial dislocation can move reversibly as the temperature changes. The irreversible variation in the SFW is usually ascribed to factors like SRO, Suzuki effect, and solute impedance effect, but no consensus has been arrived [84, 94, 118]. However, in accordance with the negative SFE, we can understand the above observations better. For example, in Cu-13 at.% Al, it was observed that dissociated dislocations generated at room temperature deformation maintain their separation widths unchanged when altering temperature between 77 K and 450K [94]. According to our previous study, this alloy has a negative SFE at room temperature [48] and a positive temperature dependence [94]. Decreasing the temperature from room temperature, the SFE becomes more negative (~-15 mJ/m$^2$ [48]), but not smaller enough (together with the repulsive force, Figure 1(b)) to overcome the lattice friction and push the partials to move, otherwise, thermally induced martensitic phase transformation is expected. When increasing temperature below $T_0$, the SFE becomes larger but still negative, there is no driving force to shrink the partial separation because both the SFE and $F_{\text{int}}$ point outwards the partial pair, thus individual partial dislocations appear locked. Only when the SFE is positive and large enough at temperatures higher than $T_0$, the SFE and the repulsive force may compete with each other (and with the friction resistance) and alter the partial separation reversibly with respect to temperature as in pure fcc metals. Note that the friction stress should decrease with increasing temperature, which also contributes to the mobility of partials at high temperatures [119, 120].

Similarly, the present work renders an alternative explanation for the large variation of the partial separation $d$ of the dislocations with the same character (same $\beta$) in HEAs [29]. Smith et al. [29] measured the partial separation of the $a/2<110>\{111\}$ 60° mixed dislocations in CrMnFeCoNi HEA and found significantly large variations in the dissociation distance, compared to the cases of pure fcc metals. It was ascribed to the multicomponent nature of the alloys which have inherent concentration inhomogeneity at nanometer scale in space, thereby, the varying local SFEs [29, 69]. In fact, in Fe-Mn steels, partial separation was also found to fluctuate dramatically [88]. In a pure fcc metal with positive SFE, there is an equilibrium state of finite partial separation as expected from the force balance between the SFE, the repulsive force (and the small lattice friction resistance) at static conditions. Large deviations in partial separations from the equilibrium distance can possibly be corrected by the SFE or repulsive force driven movements of partials. Therefore, the partial separations only slightly fluctuate around the equilibrium value. While in negative SFE materials, both the SFE and repulsive force tend to separate the partials, competing against the passive lattice friction resistance, thereby, the final separation distances become more sensitive to local shear stresses during deformation (stress history) or the image forces in TEM measurements. One can easily expect that the partials that are separated into different widths by varying local shear stresses should maintain their positions after the external force is removed. Consequently, partial dislocation separations in these alloys display greater variations.

## 4. Conclusions

We have demonstrated that the extant experimental models for SFE calculations are not valid in metastable alloys, which restrict the obtained $\gamma_{\text{isf}}^{\text{Expt.}}$ always to be positive. Therefore, the $\gamma_{\text{isf}}^{\text{Expt.}}$ values will depart from the true thermodynamic stability of the fcc phase and fail to serve as a proper indicator for the activation of the underlying deformation mechanisms. On the contrary, the SFEs calculated by DFT methods show nice correlation with the Gibbs energy difference between the fcc and hcp structures in both stable and unstable fcc metals and alloys. We argue that it is of significant importance to embrace the negative SFE for properly understanding dislocation behaviors in metastable systems. Finally, the present work calls for future development of experimental methodologies for measuring the SFE in metastable alloys and a revisit to understand the behaviors of partial dislocations under various deformation conditions in light of the negative SFE.

## Acknowledgement

H.Z. acknowledges the financial support from the National Natural Science Foundation of China (No.51871175). X.S. acknowledges the financial support from the China Scholarship Council. X.S., S.L. and L.V. thank the Swedish Research Council (VR, No. 2019-04971), the Swedish Foundation for Strategic Research, the Swedish Foundation for International Cooperation in Research and Higher Education, and the Hungarian Scientific Research Fund (research project OTKA 128229). X.A. acknowledges the financial support from Australia Research Council (DE170100053) and the Robinson Fellowship Scheme of the University of Sydney (G200726). The computations were performed on resources provided by the Swedish National Infrastructure for Computing (SNIC) at the National Supercomputer Centre in Linköping partially funded by the Swedish Research Council through grant agreement no. 2016-07213 and no. 2019-04971, and by the "H2" High Performance Cluster at Xi'an, China is acknowledged.